\documentclass[aps,prb,twocolumn,superscriptaddress,showpacs,floatfix]{revtex4}
\usepackage{epsfig}

\usepackage{amssymb}
\usepackage{amsmath}

\begin{document}
\title{The effect of a small magnetic flux on the metal-insulator transition}
\author{Lukas Jahnke}\author{Jan W. Kantelhardt}\affiliation{Institut f\"ur 
Physik, Martin-Luther-Universit\"at Halle-Wittenberg, 
 06099 Halle, Germany}\author{Richard Berkovits}
\affiliation{Minerva Center and Department of Physics, Bar-Ilan University, 
Ramat-Gan 52900, Israel}

\begin{abstract}
  We numerically show that very small magnetic flux can 
  significantly shift the metal-insulator 
  transition point in a disordered electronic system. The shift we observe
  for the 3d Anderson model obeys a power law as predicted 
  by Larkin and Khmel'nitskii (1981). We compute the exponent and find 
  good agreement with the prediction. 
  However, the power law holds only for much smaller magnetic fields than
  has been previously assumed, and is accompanied by a large prefactor,
  leading to a surprising strong dependence of the transition point on the
  applied magnetic field.
  Furthermore, we show that the critical level-spacing distribution is identical in
  the presence and absence of a magnetic field if hard-wall boundary conditions 
  are applied.  This result 
  is surprising since both cases belong to different universality classes and different
  distributions have been reported for periodic boundary conditions. 
\end{abstract}
\date{28.1.2009, version 1.0}
\maketitle

\section{Introduction}
\label{sec:introduction}

The interest in measuring small magnetic fields is driven by the possibility to
build smaller magnetic storage devices with high capacity. Much progress
in the understanding of the magnetic properties of condensed matter has been 
achieved in the last decades especially in thin layers \cite{magprop}. 
Some of the results were honored by a Nobel price in physics in 2007 for the
discovery of the giant magneto resistance by Fert and Gr\"unberg.

In this paper we want to take a different path, studying the influence of a 
magnetic field on a metal-insulator transition (Anderson transition) 
\cite{anderson} rather than on the magnetization. Already in the eighties, Larkin 
and Khmel'nitskii \cite{larkin} estimated the shift of the metal-insulator 
transition point, i.e., the critical disorder $W_c$ for small magnetic fluxes 
$\phi$ as 
\begin{equation} W_c(\phi) - W_c(0) = \Delta \phi^{\beta/\nu} \qquad {\rm with}
\quad \beta=1/2 \label{larkin}
\end{equation} 
and $\nu$ the scaling exponent of the localization length, $\lambda(W) \sim
(W-W_c)^{-\nu}$. The value $\nu \approx 1.43$ was numerically calculated for 
large and random magnetic fluxes in 3D cubic lattices \cite{slevin1997}.
The prediction (\ref{larkin}) was verified with different sophisticated 
analytical studies 
\cite{yuan91,biafore90,castellani90,wegner96,oppermann84,belitz84,shapiro84}. 
On the other hand, numerical work 
concentrated on the effect of large or random magnetic fluxes \cite{slevin1997}. 
Computation of
the shift of the mobility edge as function of the 
magnetic field shift was performed only 
once\cite{droese1998}, with large error bars and
for relatively large fluxes.  However, since Eq.~(\ref{larkin}) should be correct 
only for small fluxes a more thorough numerical analysis is needed.

Here we show that for small magnetic fluxes the shift of the critical disorder 
$W_c$ is a power law following Eq.~(\ref{larkin}) very closely. Deviations occur 
for $\phi > 0.03$, in particular for the $\phi$ values previously considered 
\cite{droese1998}. Moreover, our numerical results show
a surprisingly large prefactor $\Delta$ in Eq.~(\ref{larkin}), which leads to a very 
large shift even for small values of $\phi$. Basically, half of the total shift in
$W_c$ takes place for fluxes which are more than hundred times smaller than the ones 
previously considered. This could possibly be exploited for devices detecting 
small magnetic fields. Various experiments have demonstrated the possibility
to reproduce properties seen in the Anderson model including effects of magnetic 
fields \cite{mitin07,watanabe99,tousson87}.  

In addition to the shift of the critical disorder $W_c$ induced by $\phi>0$ a change 
in the critical level-spacing distribution with magnetic field was reported
\cite{batsch1996}. These results were obtained with periodic
boundary conditions. On the other hand, it is known that a change of boundary 
conditions also changes the critical level-spacing distribution 
\cite{braun1998,schweitzer1999}. Here we show that for hard-wall boundary 
conditions one finds {\it identical} critical level-spacing distribution functions 
in the presence and absence of a
magnetic flux, even though the magnetic field changes the 
universality class of the system. We also examine the second moment of the 
critical level spacing distributions and confirm its dependence on the boundary 
condition \cite{braun1998}. Furthermore, we examine its dependence on the 
magnetic flux for periodic and hard wall boundary conditions and find no 
changes in both cases.

The outline of the paper is as follows. In Section \ref{sec:model} we introduce
the model and also describe the finite-size fitting procedures we employed to 
determine the critical disorders $W_c$. In Section \ref{sec:phase} we discuss 
the main results presenting the phase diagram and comparing with Eq.~(\ref{larkin}).  
Section \ref{sec:clsd} is devoted to the effect of boundary conditions on the 
critical level spacing distribution and its second moment. Section \ref{sec:summary} 
gives conclusion and outlook.

\section{Model and finite-size scaling approach}
\label{sec:model}

To study metal-insulator transitions (MIT) on a 3D simple cubic lattice we consider 
the tight-binding Hamiltonian \cite{anderson},
\begin{equation} H = \sum_i \epsilon_i a^\dagger_i a_i - \sum_{( i,j )} t_{j,i} 
a^\dagger_j a_i, \label{ham} \end{equation}
where the first part represents the disordered on-site (node) potential
(homogeneous distribution $-W/2 < \epsilon_i < W/2$) and the second part
describes the transfer between neighboring sites $(i,j)$.  The transfer
probability is given by $t_{i,j} = \exp(i \varphi_{i,j})$ for neighboring
lattice points where $\varphi_{i,j}$ is the phase accumulation of the hopping
electron. A constant magnetic field ${\bf B}$ is applied by choosing phases
such that the flux $\phi$ per plaquete (unit square) in the $xy$-plane is 
in between $8 \times 10^{-4}$ and $0.25$ in units of the flux 
quantum $\varphi_0=hc/e$. The lowest value is derived from the smallest system 
size $L/a=14$ we consider here, with $a$ the lattice constant defining the 
length scale. For lower fluxes the magnetic length $L_H/a=1/(\sqrt{2 \pi \phi})$ 
would become smaller than $L$, and the electrons will not
accumulate a $2\pi$ phase shift. The largest flux we consider is given by 
the symmetry of the problem \cite{hofstadter}. The relation between the flux 
per plaquete and a given magnetic field depends on the lattice constant $a$,  
$\phi=e B a^2/h$.  We choose hard wall boundary conditions in most calculations
to avoid discontinuities in the phase accumulation of the hopping electrons 
due to the magnetic field.

To calculate the eigenvalues of 3D systems with the Hamiltonian (\ref{ham})
we use an iterative solver based on a general Davidson and Olsen algorithm
\cite{primme} where the matrix vector multiplication is performed by {\it
Intel MKL Pardiso}. We calculate six eigenvalues around $|E|=0$ for each 
configuration with linear system sizes ranging from $L/a=14$ to $L/a=40$ lattice 
points, accumulating at least $2\times 10^{5}$ eigenvalues for each size. 

It is expected that a second-order phase transition occurs for a given critical 
disorder $W_c$. For an infinite system this transition is characterized by the 
divergence of the correlation length $\lambda$. The divergence is described by 
$\lambda(W) \sim (W-W_c)^{-\nu}$ with a critical exponent $\nu$. We study this 
transition by analyzing the level-spacing distribution $P(s)$ of consecutive 
eigenvalues $E_i$, with $s=(E_i-E_{i-1})/\Delta E$ and the mean level spacing 
$\Delta E$.

\begin{figure}
  \begin{picture}(7.0,7.0)(0,0)
    \put (0,0){\includegraphics[width=7.0cm]{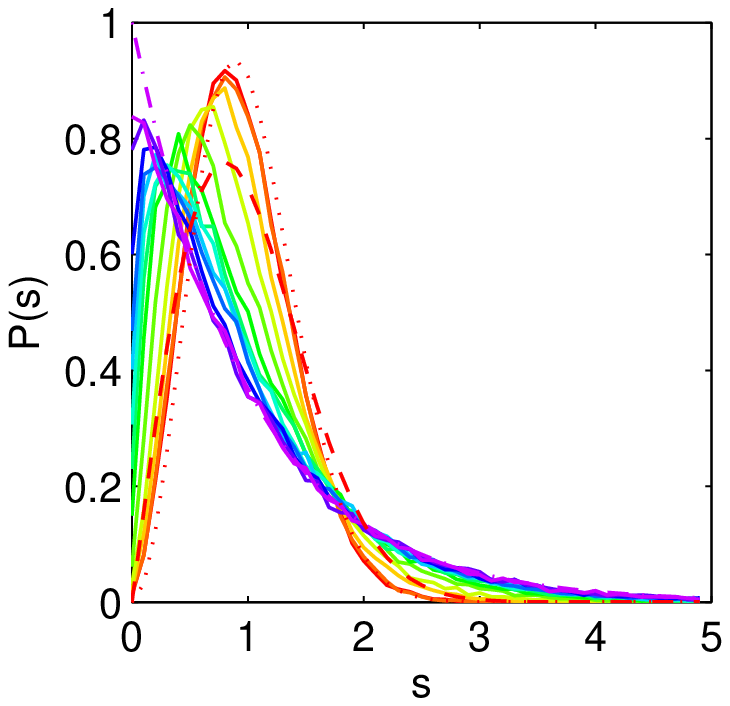}}
  \put(3.00,3.00){\includegraphics[width=3.5cm]{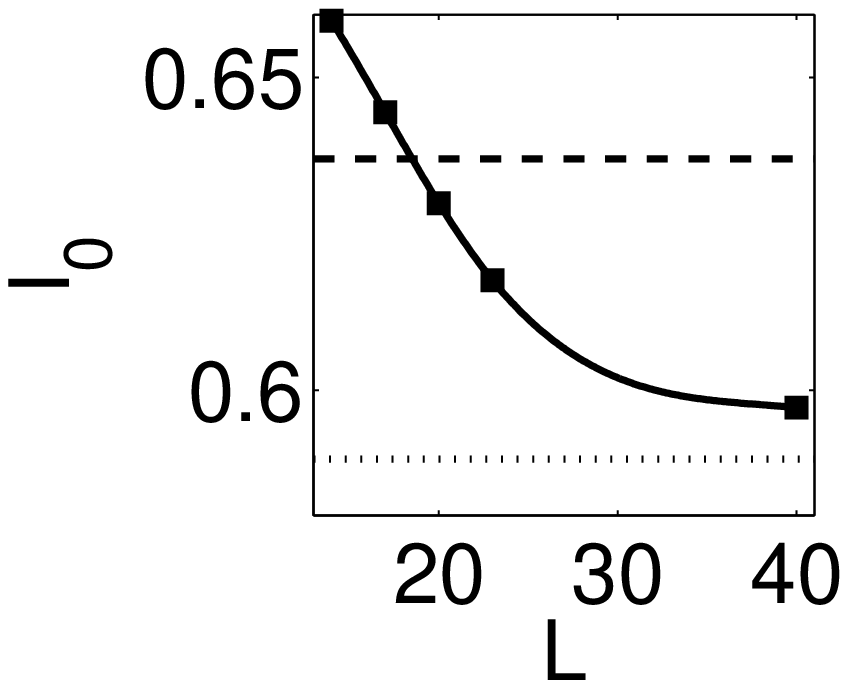}}
  \end{picture}
\caption{(Color online) The level spacing distribution $P(s)$ for a system 
of size
$L/a=40$ and magnetic flux per plaquete $\phi=10^{-3}$.
In the metallic phase (red curves) the level spacing follows the GUE distribution 
(dotted red curve) rather than the GOE distribution (dashed red curve). 
A clear transition from GUE to Poisson (dash-dotted purple curve) behavior
is observed as the disorder increases from $W=8$ (continuous red curve) to $W=23$
(continuous purple curve). Inset: System size dependence of $I_0=\langle s^2
\rangle / 2$ for $W=10$.  With increasing system size the GUE value (dotted 
line) is clearly reached asymptotically rather than the GOE value (dashed line).
\label{fig:Ps}}\end{figure}

In the limit of infinite system size the level-spacing distribution $P(s)$ of 
a disordered system corresponds to a random-matrix theory result if the 
eigenfunctions are extended. 
For systems which obey time-reversal 
symmetry the corresponding random-matrix ensemble is the orthogonal Gaussian 
ensembles (GOE), well approximated 
by the Wigner surmise, $P_{\rm GOE}(s) = (\pi/2) s \exp(-\pi s^2/4)$ (dashed 
red line in Fig.~\ref{fig:Ps}). A magnetic field breaks the time reversal 
symmetry, changing the ensemble to a unitary Gaussian ensemble (GUE), 
for which the level-spacing distribution is approximated by 
$P_{\rm GUE}(s) = (32/\pi^2) s^2 \exp(-4 s^2 / \pi)$ 
(dotted red line in Fig.~\ref{fig:Ps}). 
In contrast to the metallic phase, the uncorrelated energy levels of localized 
states are characterized by a Poisson distribution, $P_{\rm P}(s) = \exp(-s)$ 
(dashed dotted purple line in Fig.~\ref{fig:Ps}), independent of the symmetry. 

For finite systems $P(s)$ is between $P_{\rm GOE}$ 
(or $P_{\rm GUE}$ with a magnetic field) and $P_{\rm P}(s)$.  
However, it approaches one 
of them with increasing system size, remaining system-size independent only at 
the transition point, i.e. for $W=W_c$.
To determine the critical disorder we study the system-size ($L$) dependence,
of the second moment of the level spacings, $I_0= \langle s^2 \rangle/2$. 
From one-parameter finite-size scaling arguments
\cite{shklovskii1993,hofstetter92} 
we expect that $I_0$ will depend on disorder $W$ and lattice size $L$ if 
$W \ne W_{c}$, but become independent of $L$ at $W=W_c$. The
second moments of a Poissonian, GOE or GUE distribution can be calculated via
$\int_0^{\infty} s^2 P(s)$, yielding $I_{0,\rm P}=1$, $I_{0, \rm
GOE}=0.637$ (dashed line in the inset of Fig.~{\ref{fig:Ps}}) and $I_{0, \rm
GUE}=0.589$ (dotted line in the inset of Fig.~{\ref{fig:Ps}}).

Since we choose hard boundary conditions we need to take care of finite-size 
corrections due to irrelevant surface effects \cite{corrections}. Therefore we 
fit our data to a scaling form including these irrelevant surface effects 
decaying with system size as a power law, $I_0=F(\Psi L^{1/\nu}, \Xi L^y)$ 
with the critical exponent $\nu$, the relevant scaling variable $\Psi$,
the leading irrelevant variable $\Xi$ and the leading irrelevant exponent $y$. 
For finite system sizes $L$ no phase transition takes place and $F$ is a 
smooth analytical function. After expanding $F$ to first order one gets:
\begin{eqnarray} 
  I_0(\phi,W,L) &=& F_0(\Psi L^{1/\nu}) + \Xi  L^y F_1(\Psi L^{1/\nu}). 
\label{eq:IO} 
\end{eqnarray}
where
$F_0$ and $F_1$ are  analytical functions and are expanded to third
order in the analysis of the numerical results. The relevant and irrelevant 
scaling variables are expanded in power series of the dimensionless disorder 
$w=(W_c-W)/W_c$. The relevant variable is expanded to first order, $\Psi(w)=
\Psi_1 (W_c-W)/W_c$, whereas we expand to zeroth order the irrelevant 
variable, $\Xi(w)=\Xi_0$. In total there are eleven independent fit parameters 
with $W_c$ and $\nu$ being the interesting ones (see Ref. 
\cite{corrections} for the details of the procedure).  

Figure \ref{fig:Ps} shows the level-spacing distributions of a
system of size $L/a=40$ and magnetic flux $\phi=10^{-3}$ per plaquete. 
A transition can be observed from a metallic phase, where the level-spacing 
distribution is close to GUE, for $W=8$ (continuous red line), to an 
insulating phase with $P(s)$ close to Poisson, for $W=23$ (continuous 
purple line). When comparing the continuous red line with the dotted
red curve, a clear similarity can be observed. This means that the 
metallic state is similar to a random matrix GUE ($P_{\rm GUE}(s)$) and
not GOE ($P_{\rm GOE}(s)$). We do not observe a smooth
crossover between GOE and GUE for linear system sizes larger than $L_H$.
The same asymptotic behavior can also be seen in the size dependence of
the second moment of the level spacings $I_0$ for a metallic state.
For large system sizes and low disorder ($W=10$), $I_0$ reaches the GUE
value asymptotically as depicted in the inset of Fig.~\ref{fig:Ps}.  
The critical (system-size independent) level-spacing distribution occurs 
at $W_c \approx 17$ (continuous light blue line). This distribution will 
be discussed in more detail in Section \ref{sec:clsd}.

\begin{figure}
  \begin{picture}(7.0,13.0)(0,0)
    \put(0,6.5){\includegraphics[width=7.0cm]{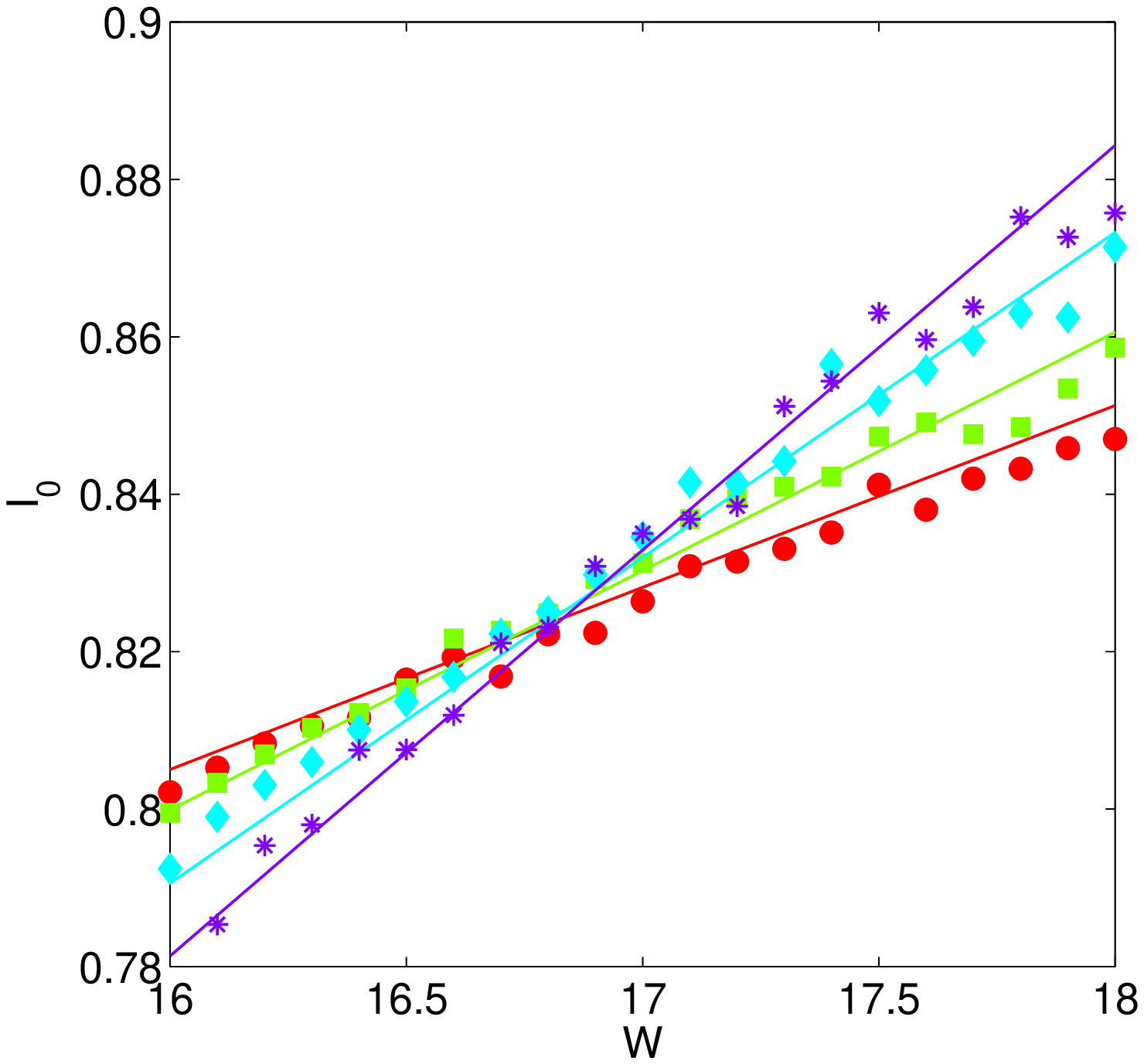}}
    \put(1.3,10.15){\includegraphics[width=2.7cm]{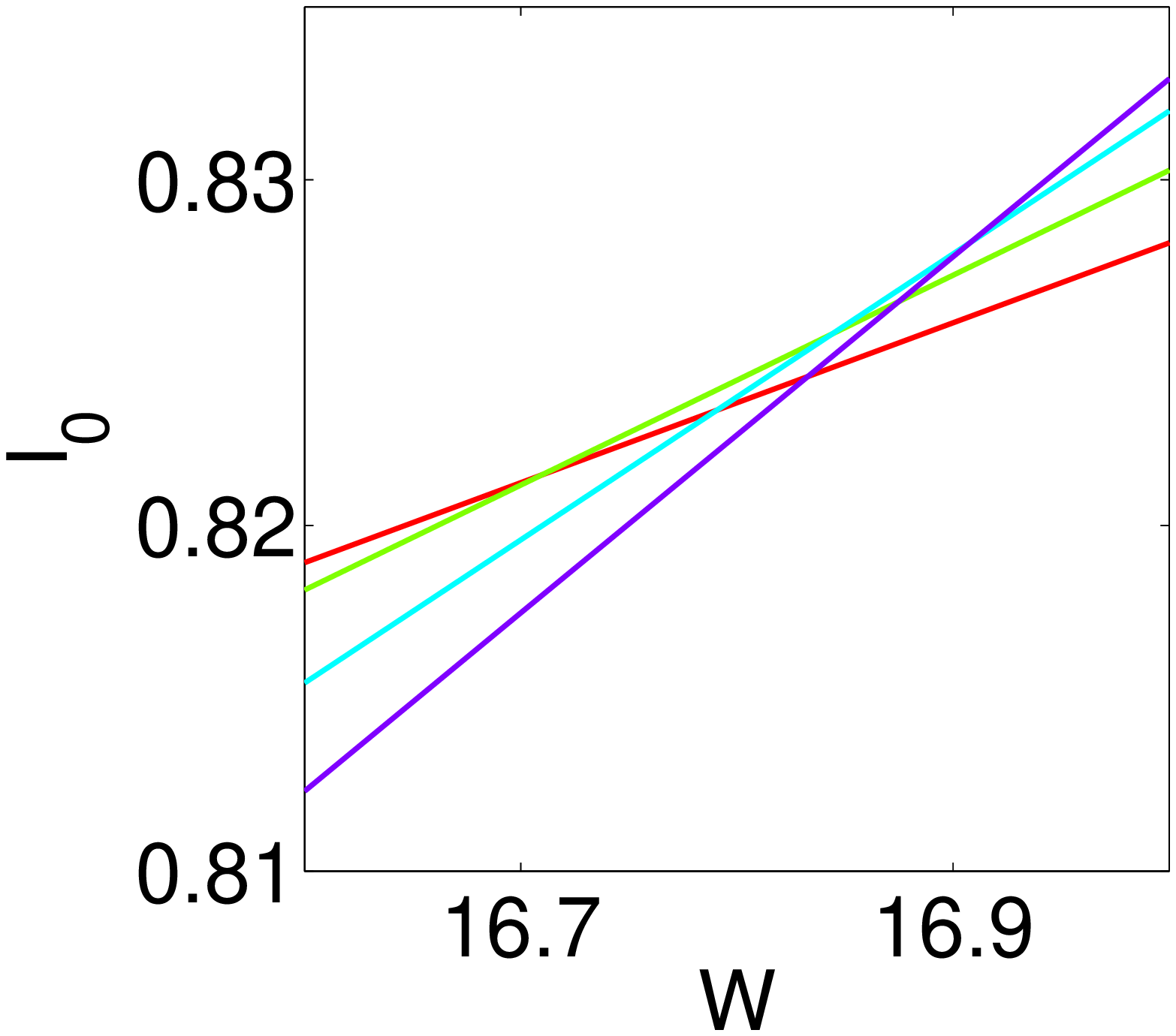}}
    \put(0.0,6.5){\parbox[t]{0.5cm}{(a)}}
    \put(0.0,0.0){\includegraphics[width=7.0cm]{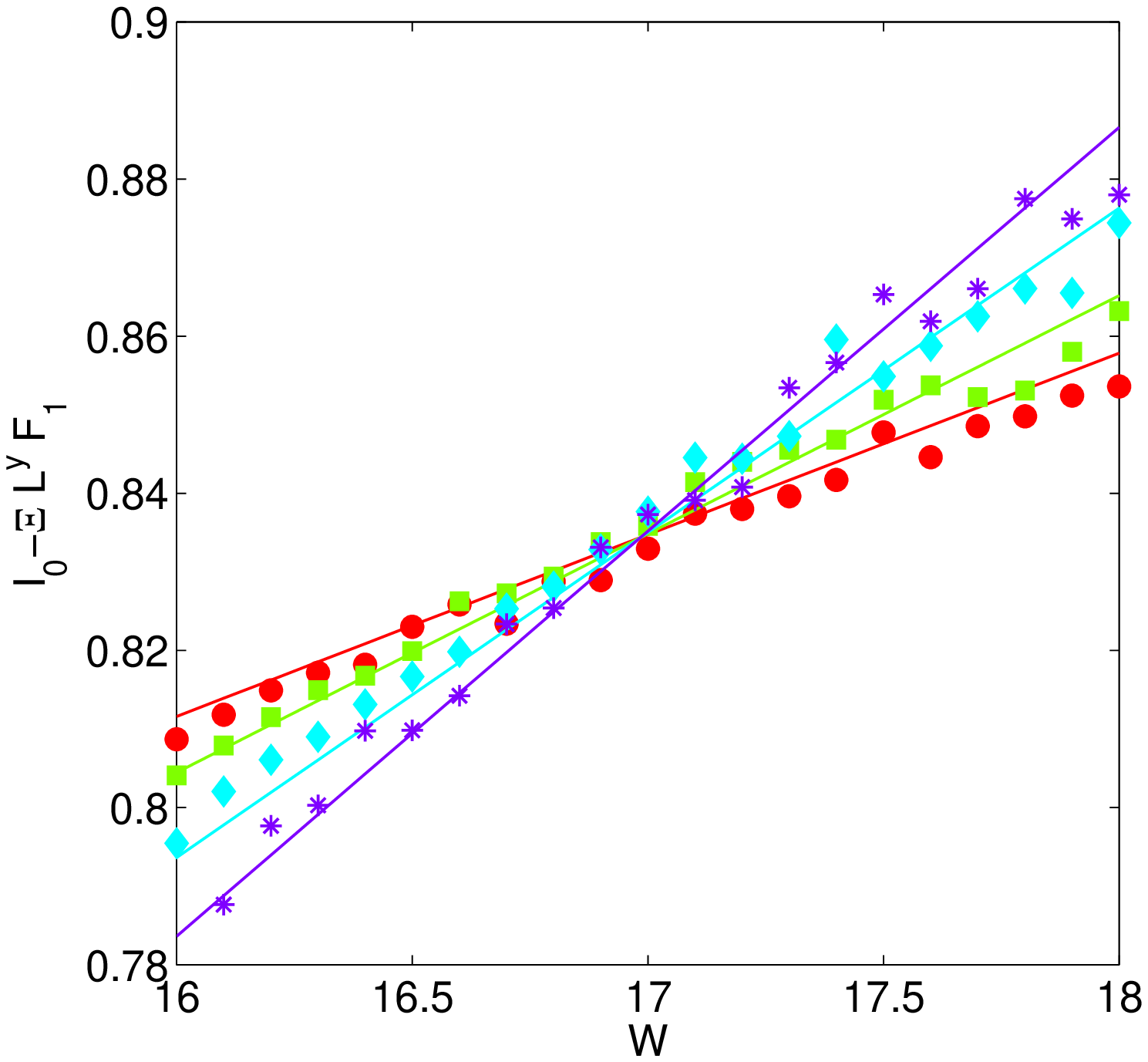}} 
    \put(1.3,3.55){\includegraphics[width=2.8cm]{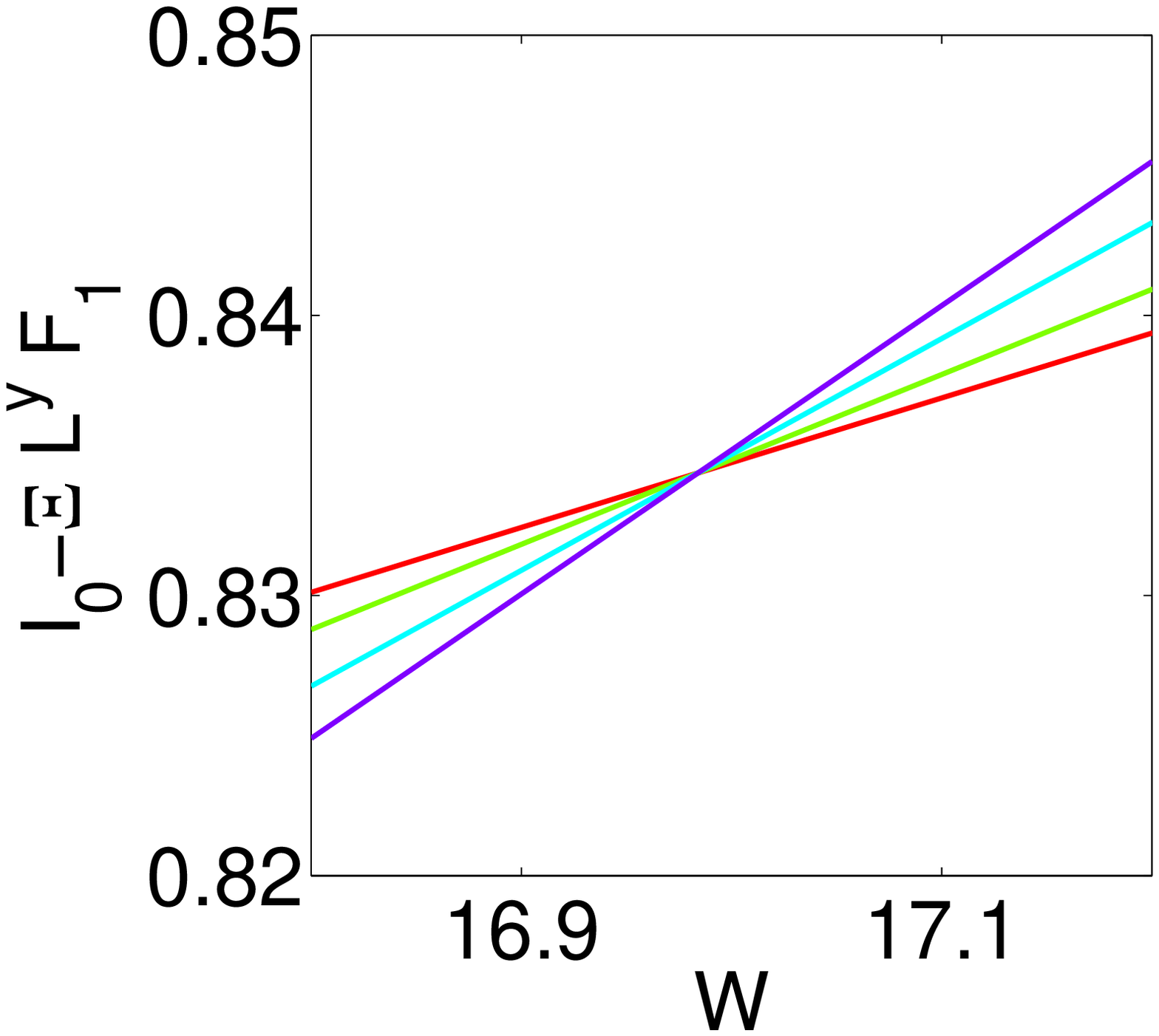}} 
    \put(0.0,0.0){\parbox[t]{0.5cm}{(b)}}
  \end{picture}
\caption{(Color online) The localization parameter $I_0=\langle s^2 \rangle/2$
versus disorder $W$ for system sizes $L/a=14$ (red circles), $L/a=20$ (green 
squares), $L/a=30$ (light blue diamonds), and $L/a=40$ (blue stars)
where a magnetic flux of $\phi=10^{-3}$ is applied.  
The lines correspond to fits of Eq.~(\ref{eq:IO}).  
(a) The symbols correspond to $I_0$. A transition from extended states for small 
$W$ to localized ones for large $W$ is seen for all sample sizes. Nevertheless,
the critical crossing point 
cannot be observed directly due to finite
size effects. Inset: the region around $W=16.8$ zoomed in; there is no 
single crosspoint.
(b) Corrected values of $I_0$ where the influence of the
irrelevant variables is subtracted. 
A clear transition can be seen at $W_{c} \approx 17$; all lines cross
at one distinct critical disorder. 
\label{fig:IO}}\end{figure}

Since the actual form of the critical level-spacing distribution is not
known we study $I_0$ to extract the critical disorder values $W_c$
and the corresponding $I_{0,c}$. In the procedure we fit our results of 
$I_0$ to Eq.~(\ref{eq:IO}) using a least-square method, i.e., a state 
of the art non-linear least-square fitting algorithm implemented in {\it 
Mathematica}.  To achieve better accuracy we do not fit the results for 
each flux $\phi$ separately but simultaneously calculate a
non-linear fit over certain ranges of fluxes. Such an approach is possible 
since $W_c$ is the only parameter in Eq.~(\ref{eq:IO}) changing significantly 
with $\phi$ for small values of $\phi$. To check the stability of the fits we compare
several fits over different ranges of flux between $\phi=8 \times 10^{-4}$ 
and $\phi=3 \times 10^{-2}$. 
For large flux ($\phi=0.1$ and $0.25$) as well as for no flux 
($\phi=0$) we perform separate fits, since other parameters than $W_c$ 
(in particularly $\nu$) change for the GOE ensemble ($\phi=0$) and for large $\phi$.

Figure \ref{fig:IO}(a) shows results of such a non-linear fit for $\phi=10^{-3}$ 
together with the simulation results for four system sizes ($L/a=14,20,30,40$). 
As expected and already indicated in the inset of Fig.~\ref{fig:Ps}, $I_0$ drops
with increasing system sizes reaching towards $I_{0,\rm GUE}$ for large system 
sizes if $W<W_c$. For $W>W_c$, on the other hand, $I_0$ increases with system 
sizes towards $I_{0, \rm P}$. At the critical disorder $W_c$, $I_0$ is 
system-size independent. The lines are the result of the simultaneous fit. 
Although it seems that all lines cross at $W \approx 16.8$ this is not the 
case as can be seen in the inset. This is due to the non-relevant variables 
$\Xi L^y F_1$ scaling with system size $L$.  
Figure \ref{fig:IO}(b) shows that subtracting the non-relevant variables leads 
to a nice crossover for $W_c \approx 17$ indicating a system-size independent 
critical value of $I_{0c} \approx 0.83$.  Taking the whole range of fluxes 
it is possible to draw a phase diagram of fluxes and critical disorders.

\section{Phase diagram and scaling behavior of the transition}
\label{sec:phase}

We have performed extensive numerical simulations for the entire range of 
magnetic fluxes \cite{hofstadter} from $\phi=0$ to $0.25$. The data is 
analyzed according to the description given in the previous section.  The 
results for the extreme cases, $\phi=0$, 0.1, and 0.25, are presented in 
Tab.~\ref{tab:resultssingle}. For $\phi=0$ we retrieve the well-known GOE 
values for $W_c$ and $\nu$ \cite{corrections}. For $\phi=0.1$ and $\phi=0.25$ 
we also reproduce previous results \cite{slevin1997,droese1998}.  
Interestingly, the critical value $I_{0,c}$ does not change upon applying the 
flux, which is also in agreement with previous work considering periodic 
boundary conditions \cite{slevin1997,droese1998,braun1998}.  

\begin{table}[t]
  \caption{Result of single fits for the interesting parameters.
  \label{tab:resultssingle}}
  \begin{ruledtabular}
  \begin{tabular}{cccc}
    $\phi$ & $0$ & $ 0.1 $ & $ 0.25 $ \\
    \noalign{\smallskip}\hline\noalign{\smallskip}
     $W_c$ & $16.59$ & $18.21$ & $18.30$  \\
     $\nu$ & $1.54$ & $ 1.42$ & $1.49$ \\
     $I_{0,c}$ & $ 0.83 $ & $0.83$ & $0.81$ \\
  \end{tabular}
\end{ruledtabular}
\end{table}

\begin{figure}[tbh]
  \begin{picture}(7.0,7.0)(0,0)
    \put(0,0){\includegraphics[width=7.0cm]{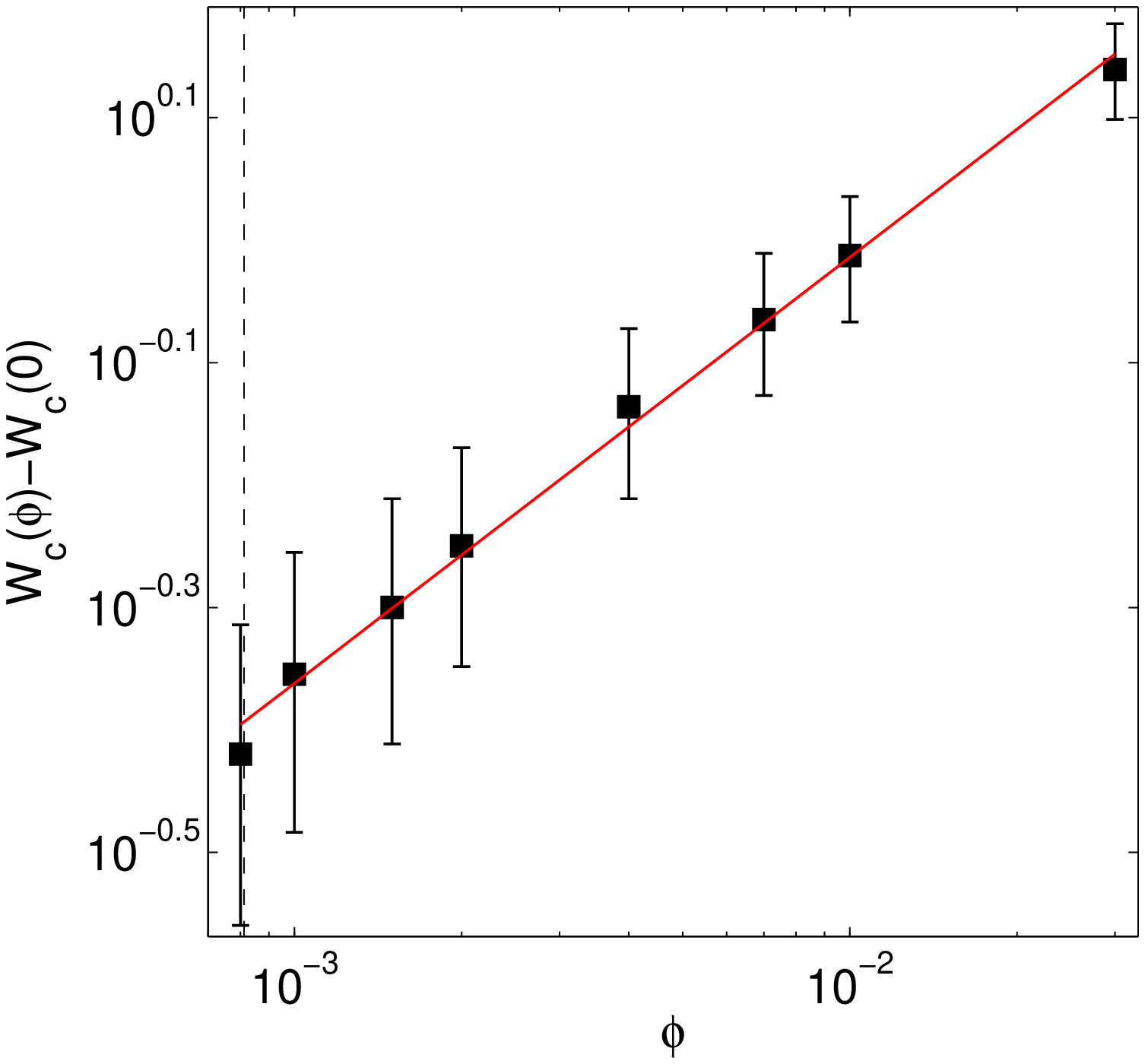}} 
    \put(3.95,0.80){\includegraphics[width=2.9cm]{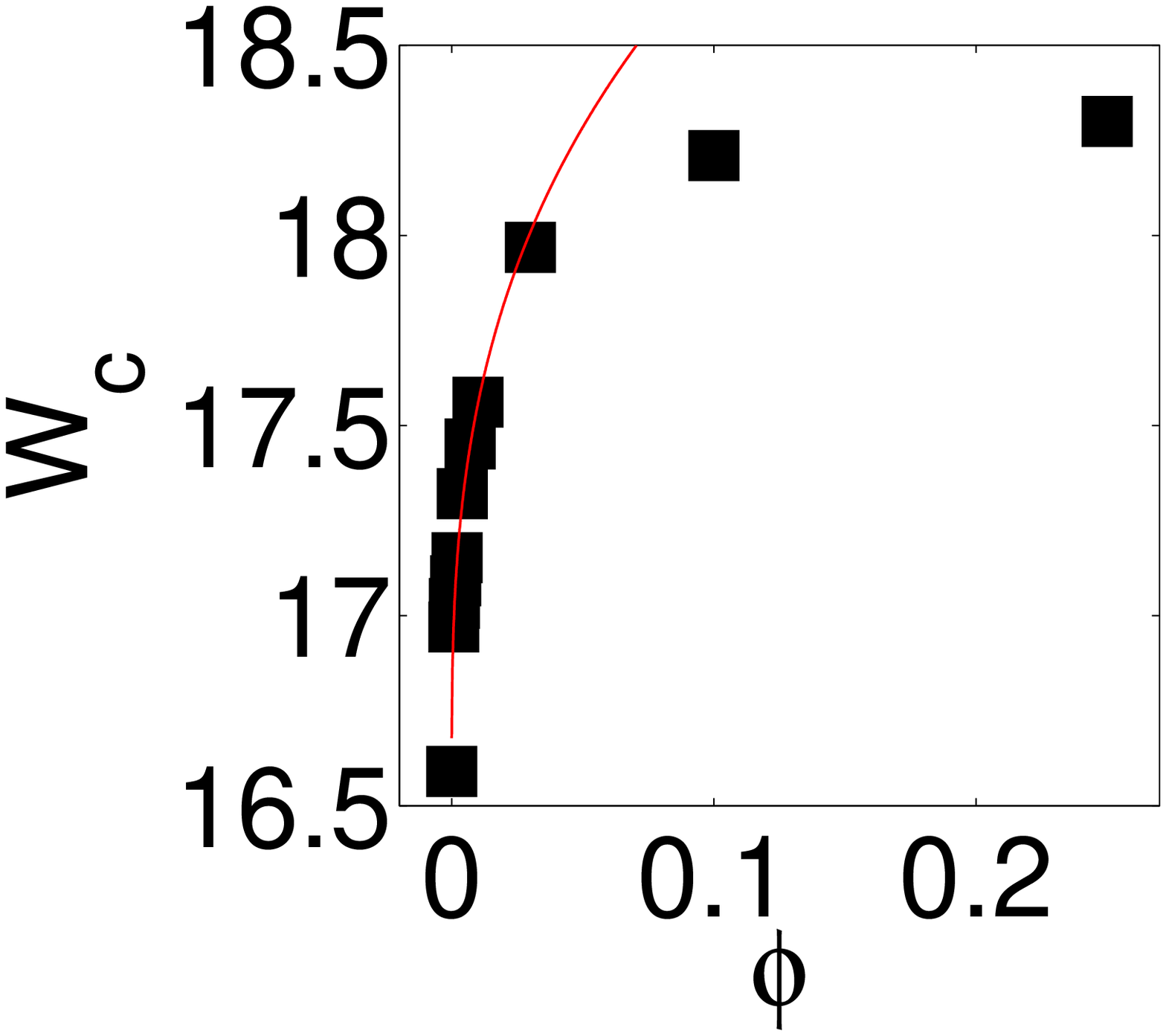}}
  \end{picture}
\caption{(Color online) The shift of the critical disorder $W_c$ as function
    of the applied magnetic field. The symbols correspond to the
    means obtained in a simultaneous non-linear fit of Eq.~(\ref{eq:IO}) with
    fluxes $\phi$ between $8 \times 10^{-4}$ and $3 \times 10^{-2}$. The error bars are the
    absolute deviation from the mean for different initial fitting parameters. 
    The red line is a linear fit of Eq.~(\ref{larkin}) with
    $\Delta=1.57$ and 
    $\beta/\nu=0.35$. Hence for $\langle \nu \rangle=1.41$ we finds $\beta=0.49$ 
    in perfect agreement with the predicted value $\beta=1/2$
    \cite{larkin,yuan91,biafore90,castellani90,wegner96,oppermann84,belitz84,shapiro84}. 
    Inset: The critical disorder $W_C$ as function of $\phi$ for larger values of the 
    magnetic flux. The squares correspond to the values of $W_c(\phi)$ depicted in
    Tab.~\ref{tab:resultssingle} and \ref{tab:results}. The red line is the power law fit used in
    in the main panel.
    $\phi=0$ and $\phi=0.25$ are symmetric axes \cite{hofstadter} and therefore the
    whole range of $\phi$ is covered. One can see the fast shift of $W_c$ for
    $\phi<3 \times 10^{-2}$, which is the region for which the scaling holds.
    \label{fig:larkin}}
\end{figure}

Figure~\ref{fig:larkin} summarizes our results for small fluxes, 
$\phi=8 \times 10^{-4}, 10^{-3},1.5 \times 10^{-3}, 2 \times 10^{-3}, 
4 \times 10^{-3}, 7 \times 10^{-3}$, $10^{-2}$, and $3 \times 10^{-2}$, 
based on a simultaneous 
fit.  The dashed line marks the flux where the magnetic length $L_H/a=
\frac{1}{\sqrt{2 \pi \phi}}$ coincides with our smallest system size 
$L/a=14$.  The values of the most important parameters $W_c$, $\nu$ and $I_{0c}$ 
are also reported in Tab.~\ref{tab:results}. They are means of the results 
achieved from different starting values in the fitting procedure; the errors 
are the absolute deviations from the means.
The result for the critical exponent at small fluxes $\nu=1.41$ 
is slightly lower but 
within the error bar range of a previous numerical study \cite{slevin1997} 
($\nu=1.43\pm 0.04$). This result 
is similar to our results for larger fluxes ($\phi=0.1$ and $0.25$).

\begin{table*}[tbh]
  \caption{Results and errors of simultaneous fit for the interesting
  parameters.   \label{tab:results}}
  \begin{ruledtabular}
    \begin{tabular}{cccccccccccc}
      & $W_c(0.0008)$ & $W_c(0.001)$ & $W_c(0.0015)$ & $W_c(0.002)$ & $W_c(0.004)$ 
      & $W_c(0.007)$ & $W_c(0.01)$ & $W_c(0.03)$ & $I_{0c}$ & $\nu$ \\
      \noalign{\smallskip}\hline\noalign{\smallskip}
      best fit & $16.97$ & $17.03$ & $ 17.09$ & $17.15$ & $17.32$
      & $17.45$      & $17.56$  & $17.97$   & $0.838$  & $1.41$   
      \\
      error    & $0.10$ & $0.11$ & $0.11$ & $0.11$ & $0.12$
      & $0.11$      & $0.11$ & $0.12$     & $0.006$  & $0.04$  \\
    \end{tabular}
  \end{ruledtabular}
\end{table*}

Our main goal , however, is to study 
the dependence of the critical disorder $W_c$ on $\phi$.  
Since the fitting approach does not use $\phi$ as a fitting parameter,
the power-law behavior shown in Fig.~\ref{fig:larkin} confirms
the prediction (\ref{larkin}) by Larkin and Khmel'nitskii \cite{larkin} 
discussed in the introduction.  Although the individual error bars for 
$W_c(\phi)$ are still significantly large, the power-law behavior for $W_c(\phi)$
comes out naturally, even though Eq.~(\ref{larkin}) is not used in our scaling
approach. The calculated exponent $\beta=0.49 \pm 0.08$ is in perfect 
agreement with the predicted exponent $\beta=1/2$.\cite{larkin,yuan91,biafore90,castellani90,wegner96,oppermann84,belitz84,shapiro84}  
The error bar 
represents the deviations we get in the exponent 
when fitting the different results of
$W_c(\phi)$ for the different initial fit variables. When different 
regimes of $\phi$ are used for the simultaneous fit, all $W_c(\phi)$ values shift 
slightly in some direction, but $\beta$ stays the same.  Therefore $\beta$ 
is more exact than the error bars in Fig.~\ref{fig:larkin} might indicate.

The inset of Fig.~\ref{fig:larkin} shows the critical disorder 
 $W_c(\phi)$ (indicated by squares) as
function of $\phi$. The red 
line is the same power-law fit as in the main panel. 
Although the prediction of 
Larkin and Khmel'nitskii were verified it is nevertheless quite surprising
how fast 
the critical disorder rises for small magnetic fluxes. 
Already at 
$\phi=0.03$ most of the shift in the critical disorder has taken place. 
For higher magnetic fluxes ($\phi>0.03$) the deviations from the
power-law scaling are seen.
We believe that such a sensitivity to small magnetic fluxes may be used as
a basis for building a magnetic sensor. 

\section{The critical level spacing distribution}
\label{sec:clsd}

In contrast to the critical disorder $W_c$ the critical level-spacing distribution
$P(s)$ is not universal, and it depends on the boundary conditions \cite{braun1998}. 
In the absence of a magnetic field \cite{braun1998} the peak of the 
distribution is shifted to 
smaller values of $s$ when the boundary conditions are successively changed 
from periodic to 
hard-wall in each of the three spatial directions.  The tails of $P(s)$ for large 
$s$ can be fitted by an exponential $P(s) \sim \exp(-\kappa s)$ with $\kappa=1.87$ 
for periodic boundary conditions \cite{batsch1996} and $\kappa=1.49$ for hard-wall 
(Dirichlet) boundary conditions \cite{schweitzer1999}.  The critical level spacing 
distribution does not only change with boundary conditions but also when a 
magnetic flux is applied. The shape shifts towards a more GUE-like shape for 
periodic boundary conditions \cite{batsch1996}. However, we are not aware of any
numerical calculation for hard-wall boundary conditions.

\begin{figure}[t]
  \begin{picture}(7.0,7)(0,0)
    \put (0,0){\includegraphics[width=7cm]{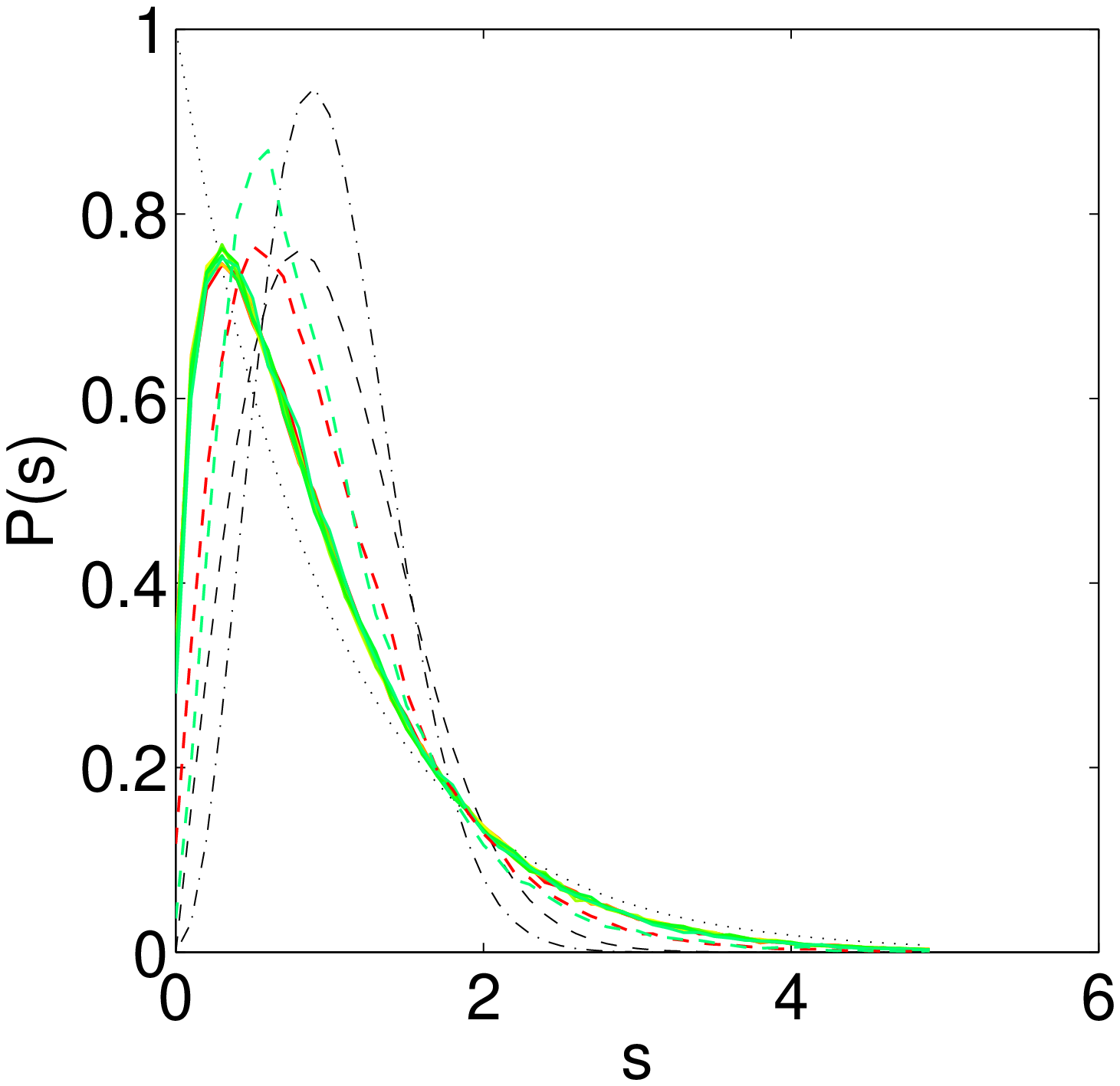}}
    \put (2.85,2.70){\includegraphics[width=3.9cm]{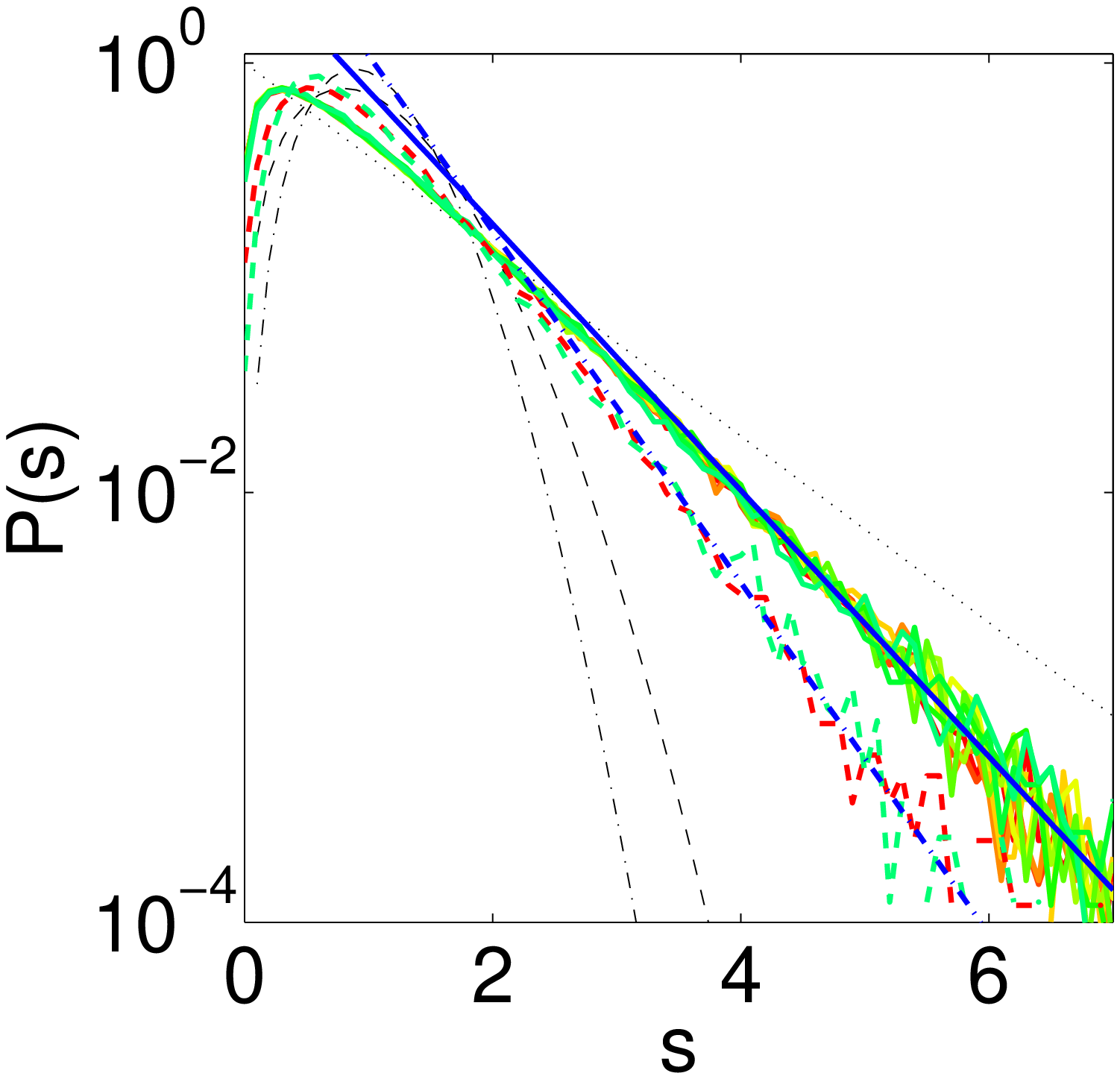}} 
  \end{picture}
\caption{(Color online) The critical level spacing distribution functions for 
periodic and hard-wall boundary conditions and different fluxes $\phi$.
For periodic boundary conditions they depend on $\phi$, where the dashed red line 
corresponds to $\phi=0$ and the dashed green line to $\phi=0.1$.  For hard-wall boundary 
conditions, there is no such dependence, as can be seen
from the behavior of the continuous lines from red to green corresponding to
$\phi=0$, $8 \times 10^{-4}$, $ 10^{-3}$, $1.5 \times 10^{-3}$, 
$2 \times 10^{-3}$, $4 \times 10^{-3}$, $ 7 \times 10^{-3}$, $ 10^{-2}$,
$3 \times 10^{-2}$, $0.1$, and $0.25$.  The limiting distributions
are also shown, black dotted 
line -- Poisson distribution, black dashed line -- GOE, black dashed dotted 
line -- GUE.  Inset: a semi-log plot depicting the behavior for large $s$. 
The solid blue lines are fits to the mean tails with $\kappa=1.42$ in the 
case of hard-wall boundary conditions and $\kappa=1.88$ for periodic 
boundary conditions, both independent of $\phi$.
\label{fig:PsWc}}
\end{figure}

Figure \ref{fig:PsWc} shows the critical level-spacing 
distribution functions $P(s)$ for Anderson models with different magnetic 
fluxes $\phi$ and periodic as well as hard-wall boundary conditions.  The
results fully confirm the previous studies discussed in the previous paragraph.
However, the critical level spacing distribution $P(s)$ is independent of $\phi$.
This new behavior of $P(S)$ for hard-wall boundary conditions is
surprising, since the universality class changes from $\phi=0$
(limit: GOE) to $\phi>0$ (limit: GUE) as can be seen by the different $\nu$
values (see Table I). In addition, the critical level distribution function
changes when periodic boundary conditions are applied. The 
tails of the critical distribution confirm to the ansatz $P(s) \sim 
\exp(-\kappa s)$ with $\kappa=1.88$ for periodic boundary conditions and
$\kappa=1.42$ for hard-wall boundary conditions independent of $\phi$.
The fits are depicted in the inset of Fig.~\ref{fig:PsWc}.

The critical second moment $I_{0,c}$ depends also on the boundary conditions.
In the literature one finds values for periodic boundary conditions,
$I_{0,c}=0.71$ \cite{braun1998,varga1995} and for hard-wall boundary 
conditions $I_{0,c}=0.81$ \cite{braun1998} for $\phi=0$.  We confirm both 
numbers numerically for $\phi=0$ and $\phi \ne 0$.  In the case of 
periodic boundary conditions the shape of $P(s)$ changes when a magnetic 
flux is introduced but the critical second moment stays the same.

\section{Summary and concluding remarks}
\label{sec:summary}
 
Our main goal has been to confirm that the shift in
the critical disorder $W_c(\phi)$ for small values of the magnetic flux $\phi$
follows the scaling behavior 
$W_c(\phi) - W_c(0) = \Delta \phi^{\beta/\nu}$ with $\beta=1/2$ as
predicted by Larkin and Khmel'nitskii \cite{larkin}. After a careful
numerical study we are able to confirm this prediction,
however, the prefactor
$\Delta$ is much larger than expected from previous numerical studies\cite{droese1998}.
Thus, the scaling holds only for a narrow range of small fluxes up to
$\phi \approx 0.03$ resulting in the fact that very weak fluxes have a strong 
effect on the 
resistivity of a sample at low temperature.  
This type of behavior naturally leads to the observation that the effect 
might be used as the basis of 
an extremely sensitive low-temperature sensing device.

In addition we have further considered the effects of boundary conditions and magnetic 
fields on the details of the metal-insulator transition of a 
3D Anderson model. The results could be summarized 
as follows: the critical disorder $W_c$ is independent of the boundary 
conditions although it depends on $\phi$. The opposite is true for the critical second
moment $I_{0,c}$ of the level spacing distribution $P(s)$, which is independent 
of $\phi$ but changes with boundary conditions. The flux dependence
of the critical level spacing $P(s)$ is sensitive to the boundary conditions.  

{\it Acknowledgment:} This work has been supported by the Minerva Foundation, the 
Israel Science Foundation (Grant 569/07), and the Deutsche Forschungsgemeinschaft 
(DFG, within SFB 418).


\begin{thebibliography}{99}

\bibitem{magprop}  {\it Handbook of Magnetism and Advanced Magnetic
  Materials}, edited by H. Kronm\"uller and S. Parkin, (John Wiley \&
  Sons, 2007).
\bibitem{anderson} P. W. Anderson, Phys. Rev. {\bf 109}, 1492 (1958). 
\bibitem{larkin} D. E. Khmel'nitskii and A. I. Larkin, Solid State Commun.
  {\bf 39}, 1069 (1981).
\bibitem{slevin1997} K. Slevin and T. Ohtsuki, Phys. Rev. Lett. {\bf 78}, 4083
  (1997); T. Kawarabayashi, B. Kramer, and T. Ohtsuki, Phys. Rev. B {\bf 57},
  11842 (1998); T. Ohtsuki, K. Slevin, and T. Kawarabayashi, Ann. Phys. (Leipzig)
  {\bf 8}, 655 (1999).
\bibitem{yuan91} JY. Yuan and T. R. Kirkpatrick, J. Stat. Phys. {\bf 64},
  309 (1991).
\bibitem{biafore90} M. Biafore, C. Castellani, and G. Kotliar, Nucl. Phys. B
  {\bf 340}, 617 (1990).
\bibitem{castellani90} C. Castellani and G. Kotliar, Physica A {\bf 167},
  294 (1990).
\bibitem{wegner96} F. J. Wegner, Nucl. Phys. B {\bf 270}, 1 (1996).
\bibitem{oppermann84} R. Oppermann, J. Phys. Lett. (Paris) {\bf 45}, 1161
  (1984).
\bibitem{belitz84} D. Belitz, Solid State Commun. {\bf 52}, 989 (1984).
\bibitem{shapiro84} B. Shapiro, Philos. Mag. B  {\bf 50}, 241 (1984).
\bibitem{droese1998} T. Dr\"ose, M. Batsch, I. K. Zharekeshev, and B. Kramer,
  Phys. Rev. B {\bf 57}, 38 (1998).
\bibitem{mitin07} V. F. Mitin, V. K. Dugaev, and G. G. Ihas, App. Phys.
  Lett. {\bf 91}, 202107 (2007).
\bibitem{watanabe99} M. Watanabe, K. M. Itoh, Y. Ootuka, and E. E.
  Haller, Phys. Rev. B {\bf 60}, 15818 (1999).
\bibitem{tousson87} E. Tousson, V. Volterra, E. P.
  Rubenstein, R. Rosenbaum, and Z. Ovadyahu, Philos. Mag. B {\bf 56}, 875
  (1987).
\bibitem{batsch1996} M. Batsch, L. Schweitzer, I. Kh. Zharekeshev, and B.
  Kramer, Phys. Rev. Lett. {\bf 77}, 1552 (1996).
\bibitem{braun1998} D. Braun, G. Montambaux, and M. Pascaud, Phys. Rev. Lett.
  {\bf 81}, 1062 (1998).
\bibitem{schweitzer1999} L. Schweitzer and H. Potempa, Physica A {\bf 266},
  486 (1999).
\bibitem{hofstadter} D. R. Hofstadter, Phys. Rev. B {\bf 14}, 2239 (1976).
\bibitem{shklovskii1993} B. I. Shklovskii, B. Shapiro, B. R. Sears, P.
  Lambrianides, and H. B. Shore, Phys. Rev. B. {\bf 47}, 11487 (1993).
\bibitem{hofstetter92}
  E. Hofstetter and M. Schreiber, Phys. Rev. B {\bf 48},
  16979 (1993); {\it ibid} Phys. Rev. B {\bf 49}, 14726 (1994).
\bibitem{corrections} K. Slevin and T. Ohtsuki, Phys. Rev. Lett. {\bf 82}, 382  (1999); K. Slevin, T. Ohtsuki, and T. Kawarabayashi, {\it ibid} {\bf 84}, 3915 (2000).
\bibitem{primme} A. Stathopoulos, SIAM J. Sci. Comput. {\bf 29}, 481 (2007);
  A. Stathopoulus and J. R. McCombs, {\it ibid} {\bf 29}, 2162
  (2007).
\bibitem{varga1995} I. Varga, E. Hofstetter, M. Schreiber, and J. Pipek, Phys.
  Rev. B {\bf 52}, 7783 (1995).

\end{thebibliography}
\end{document}